\documentclass{iaus}
\usepackage{graphics}

\title[Optical recombination lines] 
{Optical recombination lines as probes of conditions in planetary nebulae}

\author[X.-W. Liu]   
{Xiao-wei Liu}

\affiliation{Department of Astronomy, Peking University,
Beijing 100871, P. R. China \break email: liuxw@bac.pku.edu.cn}

\pubyear{2006}
\volume{234}  
\pagerange{119--126}
\date{?? and in revised form ??}
\jname{Proceedings Title IAU Symposium}
\editors{A.C. Editor, B.D. Editor \& C.E. Editor, eds.}
\begin{document}

\maketitle

\begin{abstract} Since the last IAU symposium on planetary nebulae (PNe),
several deep spectroscopic surveys of the relatively faint optical
recombination lines (ORLs) emitted by heavy element ions in PNe and H~{\sc ii}
regions have been completed. New diagnostic tools have been developed thanks to
progress in the calculations of basic atomic data.  Together, they have led to
a better understanding of the physical conditions under which the various types
of emission lines arise. The studies have strengthened the previous conjecture
that nebulae contain another component of cold, high metallicity gas, which is
too cool to excite any significant optical or UV CELs and is thus invisible via
such lines. The existence of such a plasma component in PNe and possibly also
in H~{\sc ii} regions provides a natural solution to the long-standing problem
in nebular astrophysics, i.e. the dichotomy of nebular plasma diagnostics and
abundance determinations using ORLs and continua on the one hand and
collisionally excited lines (CELs) on the other.  
\end{abstract}

\firstsection 
\section{The dichotomy between ORLs and CELs}\label{sec:intro}

Observations and analyses of emission line nebulae, PNe and H~{\sc ii} regions,
are widely used to obtain elemental abundances in our own Galaxy and beyond. A
detailed knowledge of the physical conditions under which the lines arise and a
full understanding of their excitation mechanisms are of paramount importance
for the reliability and accuracy of the results, and consequently for the study
of stellar nucleosynthesis and the chemical evolution of galaxies. However,
observations in the last decade of the relatively faint ORLs from heavy element
ions have yielded elemental abundances that are consistently higher than values
derived from the traditional method based on strong CELs. The discrepancy
between ORL and CEL abundances first drew attention and became an issue of
dispute ever since IUE measurements in the early 1980s of the ultraviolet
C~{\sc iii}] $\lambda\lambda$1907,1909 CELs yielded systematically lower
C$^{2+}$/H$^+$ abundance ratios than values derived from the C~{\sc ii}
$\lambda$4267 ORL.  Similarly, in another major dilemma, it is also found that
$T_{\rm e}$'s deduced from the Balmer discontinuity of H~{\sc i}
recombination spectra are systematically lower than those derived from the
[O~{\sc iii}] forbidden line ratio, a discrepancy first reported by
Peimbert (1971).  For most PNe, the ORL/CEL abundance discrepancy factors
(adf's) typically lie in the range 1.6--3.2, but with a significant tail
extending to much higher values.  Liu et al. (1995) and Luo et al.  (2001)
found an adf of $\sim 5$ for NGC 7009, while Liu et al.  (2000) derived an adf
of $\sim 10$ for NGC 6153 and Liu et al. (2001) obtained adf's of $\sim 6$ and
20 for Bulge PNe M~2-36 and M~1-42, respectively. The highest adf recorded
hitherto for a PN is $\sim 71$ for Hf\,2-2 (Liu 2003; Liu et al. 2006).  

\section{Interpretation}

It is now widely accepted that both types of the above-mentioned discrepancies
are real rather than caused by e.g. errors in atomic data or observational
uncertainties. The discrepancies are traditionally interpreted in terms of
large $T_{\rm e}$ fluctuations (Peimbert 1967) and/or $N_{\rm e}$
inhomogeneities (Rubin 1989; Viegas \& Clegg 1994). Yet there is no direct
measurement pointing to the existence of large $T_{\rm e}$ fluctuations in PNe
and physical causes that might lead to them remain unclear.  More importantly,
the paradigm fails to explain the wide range of data extending from the UV to
the IR that are available now (Liu 2003, 2005; but see the review by M.
Peimbert in the volume for a different view):

\begin{enumerate}

\item The value of the adf is found to be uncorrelated with the excitation
energy $E_{\rm ex}$ or the critical density $N_{\rm cr}$ of the CEL involved,
as one would expect if $T_{\rm e}$ fluctuations and/or $N_{\rm e}$
inhomogeneities are the root cause;

\item IR fine-structure lines such as [O~{\sc iii}] 52- and 88-$\mu$m, which
have $E_{\rm ex} \lesssim 1000$\,K and are therefore insensitive to $T_{\rm e}$
and $T_{\rm e}$ fluctuations, yield ionic abundances comparable to UV/optical
CELs. This is true for all PNe analyzed hitherto, including those of densities
lower than the critical densities of the IR fine-structure lines; 

\item Values of the adf larger than 5 are difficult to explain by such effects;

\item The $T_{\rm e}$ fluctuation scenario implicitly requires the nebular
heavy element abundances to be the higher ORL values.  Yet such abundances are
often much higher than the solar value making it very difficult to reconcile
with the current theory of stellar evolution for low- and intermediate-mass
stars.

\item Imaging and spatially resolved spectroscopic observations have yielded
small values for $t^2$, Peimbert's parameter of $T_{\rm e}$ fluctuations (e.g.
Rubin et al. 2002 for NGC 7009).

\end{enumerate}

An alternative interpretation was presented by Liu et al. (2000). In their
analysis of NGC\,6153, Liu et al. concluded that the nebula must contain
another previously unknown component of ionized gas, cold and highly enriched
in heavy elements. The conjecture was strongly supported by the later discovery
of very low Balmer jump $T_{\rm e}$'s in two PNe of large adf's, 3680~K in
M\,1-42 (adf = 22; Liu et al. 2001) and 900~K in Hf\,2-2 (adf = 71; Liu 2003;
Liu et al. 2006). Both PNe have a typical [O~{\sc iii}] forbidden line
temperature of $\sim 9000$~K.  Further evidence pointing to the presence of a
cold component of metal-rich plasma was provided by the analyses of He~{\sc i}
and O~{\sc ii} ORLs as well as by detailed photoionization modeling that
incorporates H-deficient inclusions embedded in a nebula of ``normal''
composition (c.f. Liu 2003, 2005 and references therein). 

In what follows, I will summarize progress achieved on this important problem
since the last IAU symposium on PNe.  Additional discussions on the topic can
be found in this volume in articles by Peimbert, by Barlow, Liu \& Hales and by
Tsamis et al.   

\section{ORL spectroscopic surveys}

\begin{figure}
\centering
\resizebox{6.0cm}{!}{\includegraphics{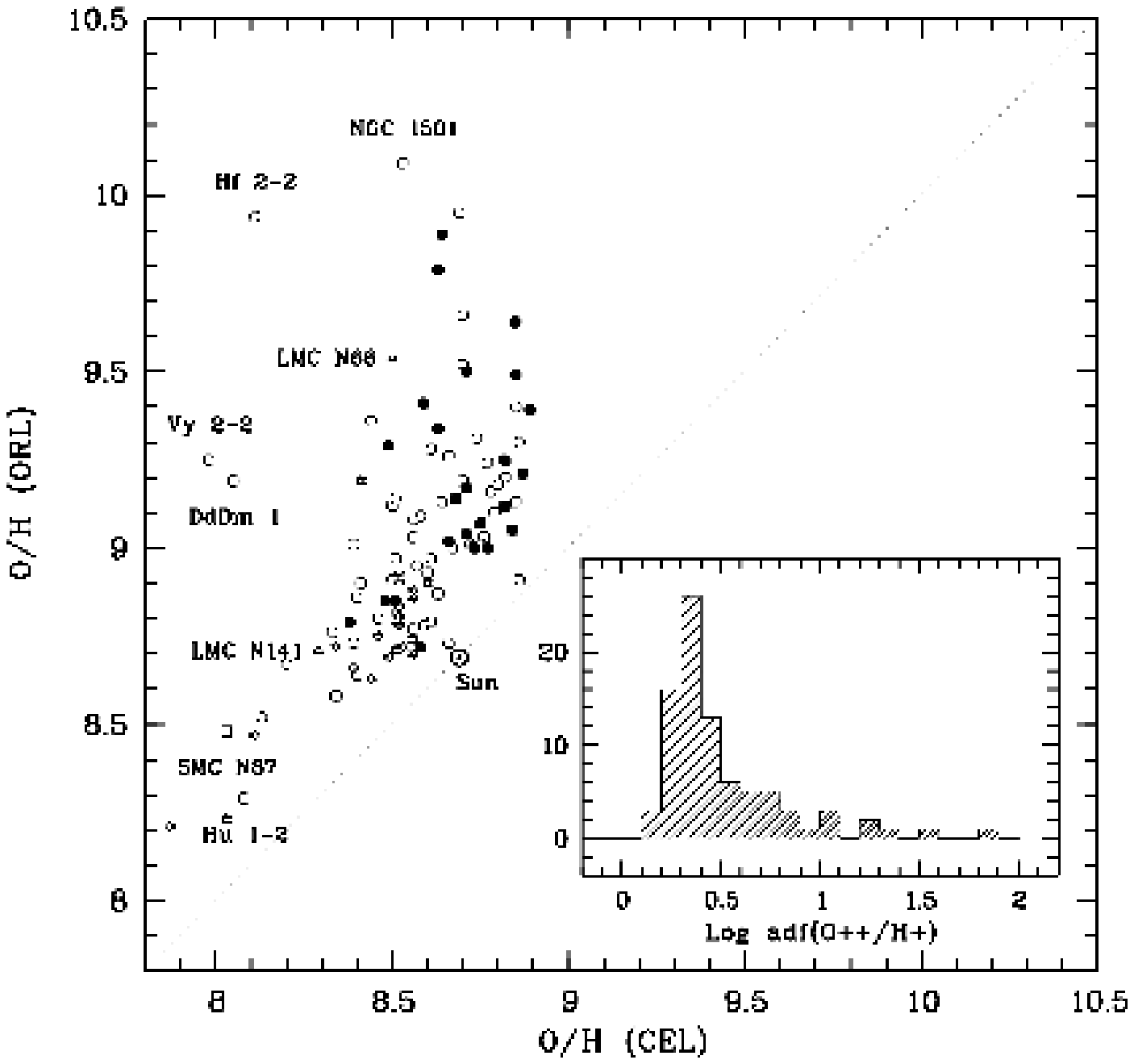} }
\resizebox{6.0cm}{!}{\includegraphics{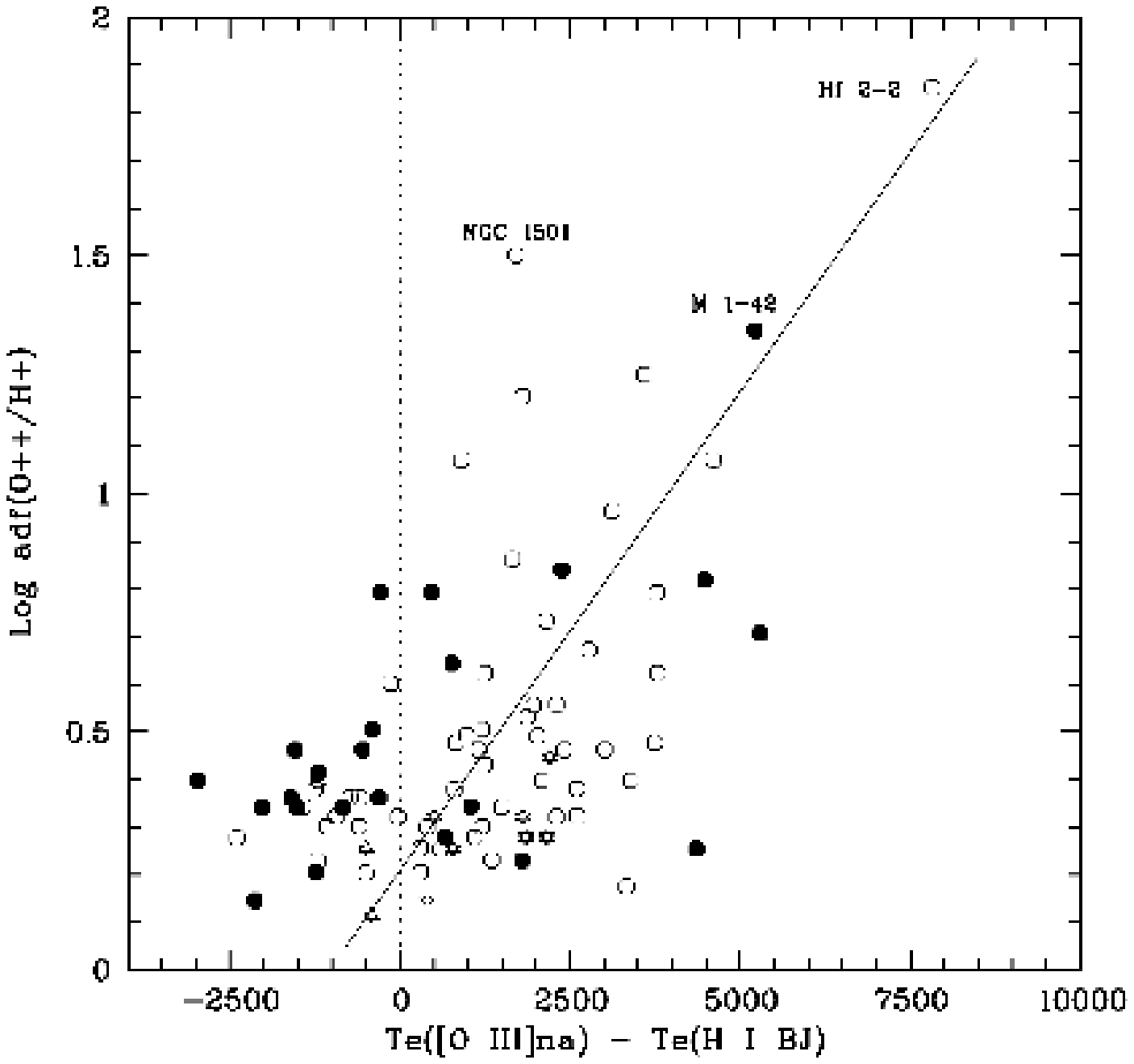} }
\caption[]{{\it Left panel:} Comparison of O/H deduced from O~{\sc ii} ORLs and
from optical [O~{\sc iii}] CELs. The insert shows a histogram of the ORL/CEL
abundance discrepancy factor (adf) of O$^{2+}$/H$^+$ for 86 PNe. {\it Right
panel:} $\log$\,adf(O$^{2+}$/H$^+$) plotted against the difference between
$T_{\rm e}$'s deduced from the [O~{\sc iii}] optical forbidden lines and 
from the H~{\sc i} Balmer jump.  The solid line shows a linear fit obtained
by Liu et al. (2001).  Open circles: Galactic disk PNe; Filled circles:
Galactic Bulge PNe; Triangles: Magellanic PNe; Stars: H~{\sc ii} regions.}
\end{figure}

Several deep optical spectroscopic surveys, allowing detailed nebular plasma
diagnostics and abundance analyses using the relatively weak hydrogen and
helium recombination continua and ORLs from heavy element ions, have been
carried out and published for several dozens of Galactic disk and Bulge PNe and
for a number of Galactic and extragalactic H~{\sc ii} regions. Tsamis et al.
(2003b, 2004) studied 12 Galactic and 3 Magellanic PNe. Liu et al. (2004ab)
analyzed 12 Galactic PNe.  Robertson-Tessi \& Garnett (2005) and Wesson et al.
(2005) surveyed 6 and 23 Galactic PNe, respectively. Finally, Wang \& Liu
(2006, in preparation; c.f. their contribution in this volume) observed 25
Galactic Bulge and 6 disk PNe. Detailed studies for individual nebulae since
2001 were presented by Liu et al. (2001; M\,1-42 and M\,2-36), Garnett \&
Dinerstein (2001a; NGC\,6720), Zhang \& Liu (2003; M\,2-24), Shen, Liu \&
Danziger (2003; Me\,1-1), Wesson, Liu \& Barlow (2003; Abell\,30), Ruiz et al.
(2003; NGC\,5307), Peimbert et al.  (2004; NGC\,5315), Ercolano et al.  (2004;
NGC\,1501), Wesson \& Liu (2004; NGC\,6543), Sharpee et al. (2004; IC\,418),
Zhang et al. (2005; NGC\,7027) and Liu et al. (2006; Hf\,2-2). In total, about
90 PNe have been studied using ORLs. For H~{\sc ii} regions, observations were
presented by Esteban et al.  (2002) for 4 extragalactic giant H~{\sc ii}
regions (NGC\,604 in M\,33, NGC\,5461 and 5471 in M\,101 and NGC\,2363), Tsamis
et al.  (2003a) for 3 Galactic (M\,42, M\,17 and NGC\,3576) and 3 Magellanic
(30 Doradus, LMC\,N\,11B and SMC\,N\,66) nebulae, Peimbert (2003) for 30
Doradus, Esteban et al. (2004) for M\,42 and Garc\'{i}a-Rojas et al. (2004,
2005 and 2006) for NGC\,3576, S\,311, M\,16, M\,20 and NGC\,3603.

The left panel of Fig.\,1 compares O/H deduced from ORLs and from CELs for PNe
and H~{\sc ii} regions. For {\em all}\, nebulae, ORL abundances are higher than
the corresponding CEL values. In addition, except for a few halo PNe, O/H
abundances deduced from ORLs are higher than the solar value of 8.69 (Lodders
2003) for almost all Galactic PNe, by up to a factor of 25 (1.4\,dex) in the
most extreme case. The insert shows a histogram of $\log$\,adf(O$^{2+}$/H$^+$)
for 86 PNe.  The distribution peaks at 0.35\,dex, or a factor of 2.
Approximately 10\% and 20\% of PNe have adf's larger than 10 and 5,
respectively. The right panel of Fig.\,1 plots $\log$\,adf(O$^{2+}$/H$^+$)
against $T_{\rm e}$([O~{\sc iii}])$-T_{\rm e}$(BJ), the difference between
$T_{\rm e}$'s derived from the [O~{\sc iii}] forbidden lines and from
the Balmer jump of the H~{\sc i} recombination spectrum. As noted by Liu et al.
(2001) there is a positive correlation between the two quantities.  For
Hf\,2-2, the most extreme PNe found so far, adf = 71 and $T_{\rm e}$(BJ)$ =
900$~K, compared to $T_{\rm e}$([O~{\sc iii}])$ = 8820$~K (Liu et al. 2006).

\begin{figure}
\centering
\resizebox{5.4cm}{!}{\includegraphics{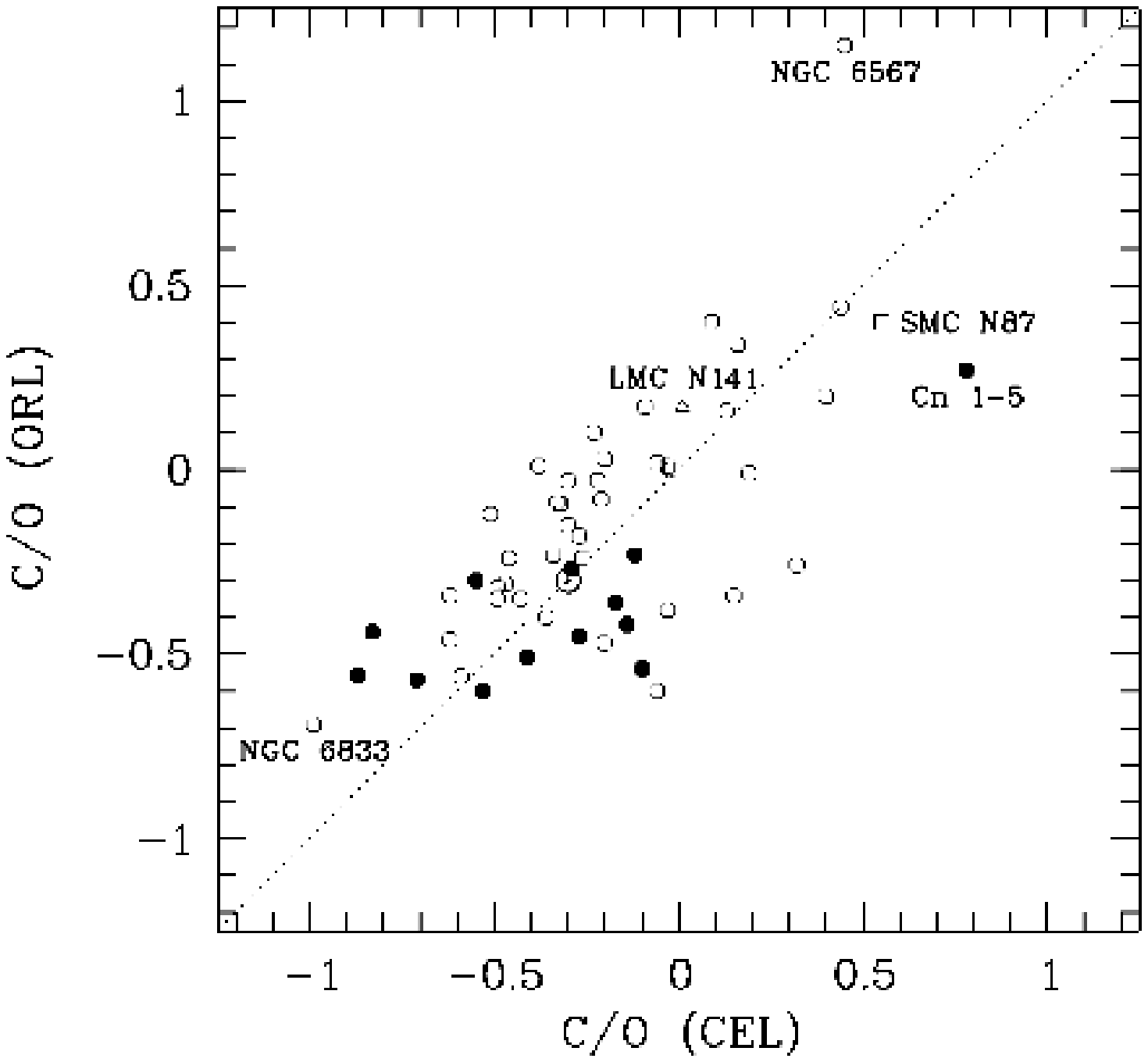}}
\resizebox{5.3cm}{!}{\includegraphics{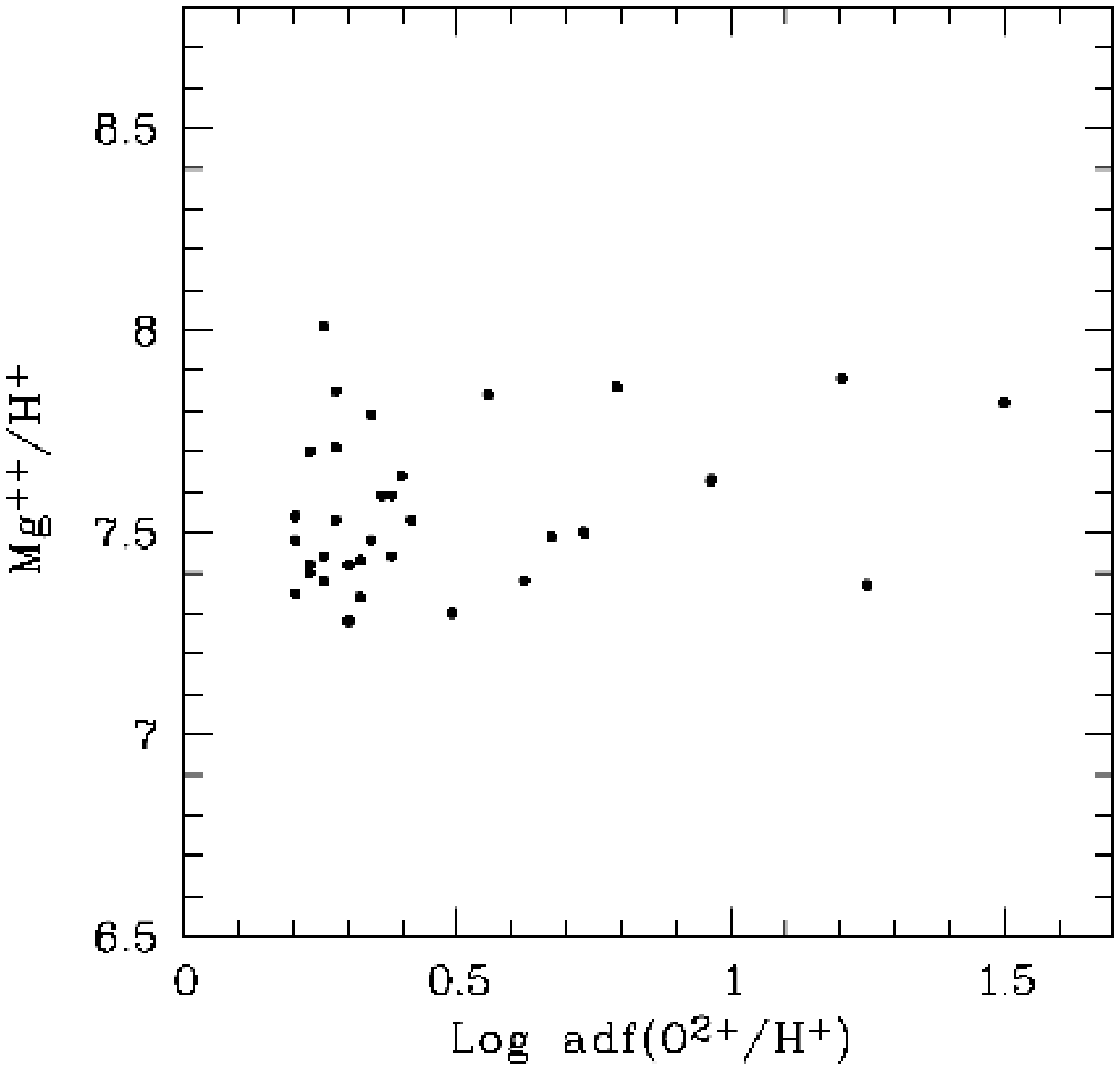}}
\caption[]{{\it Left panel:} Comparison of C/O ratios derived from ORLs and
from CELs. Open circles: Galactic disk PNe; Filled circles: Galactic Bulge PNe.
For the 39 disk and 13 Bulge PNe, ORL analysis yields average C/O ratios of
$-0.11\pm 0.05$ and $-0.38\pm 0.07$, respectively. For CEL analysis, the
corresponding values are $-0.19\pm 0.05$ and $-0.32\pm 0.12$, respectively.
For the whole sample, ORL analysis yields average C/O ratios of $-0.15\pm 0.04$
for 57 disk PNe and $-0.36\pm 0.0$ for 24 Bulge PNe, respectively.  {\it Right
panel:} Mg$^{++}$/H$^+ \approx $ Mg/H plotted against
$\log\,$adf(O$^{++}$/H$^+$). The 33 PNe shown yield an average logarithmic
Mg$^{++}$/H$^+$ ratio of $7.56\pm 0.03$ on a scale where H = 12,
almost identical to the solar value Mg/H = 7.55 (dotted line).}
\end{figure}

Large, evolved, low surface brightness, low excitation PNe tend to have large
adf's (Garnett \& Dinerstein 2001b; Tsamis et al. 2004; Liu et al.  2004b). For
a given PN, adf increases towards the centre (Garnett \& Dinerstein 2001a; Liu
et al. 2000; Luo \& Liu 2003; Liu et al. 2006). Similar patterns were observed
earlier in several PNe in the case of C$^{2+}$/H$^+$ (e.g. Barker 1987). It
seems that as the main nebula expands and decreases in surface brightness, cold
H-deficient inclusions become dominant in emitting ORLs.
 
For all abundant second-row elements studied, X/H (X = C, N, O and Ne) deduced
from ORLs are all enhanced compared to the CEL values by about the same amount
for a given nebula. Abundance ratios, such as C/O, N/O and Ne/O deduced from
two types of lines are therefore comparable. The left panel of Fig.\,2 shows
that C/O ratios deduced from ORLs and from CELs agree well.  The agreement is
less satisfactory for N/O and Ne/O -- ORLs tend to yield higher values than
CELs, by 0.2 and 0.4\,dex, respectively. The discrepancies are probably caused
by uncertainties in the effective recombination coefficients for N~{\sc ii} and
Ne~{\sc ii}.  We still lack high quality calculations of effective
recombination coefficients for the 3d -- 4f transitions. Note that Fig.\,2
shows that Bulge PNe have lower C/O ratios than disk ones by approximately
0.2\,dex.

In contrast, for the third-row element Mg, values of Mg$^{2+}$/H$^+$, which to
a good approximation equals Mg/H, deduced from the Mg~{\sc ii} 3d--4f
$\lambda$4481 ORL, are nearly constant for a wide range of adf and are almost
identical to the solar value (Barlow et al.  2003; Liu et al.  2004b).  As
pointed out by Barlow et al.  (2003), depletion of Mg onto dust grains is
unlikely significant for ionized gaseous nebulae.  Data on Mg abundances from
CELs are scarce, due to the lack of lines in the optical. For a few high
excitation PNe for which IR observations of the [Mg~{\sc iv}] 4.4$\mu$m and
[Mg~{\sc v}] 5.6$\mu$m lines are available (
Liu et al.  2004b; Zhang et al. 2005b), the data appear to imply adf(Mg/H)
$\sim 1$. It seems that the large ORL/CEL abundance discrepancies observed for
second-row elements are not present for magnesium. The result poses a serious
constraint on the possible origins of H-deficient inclusions that are
postulated to exist in PNe and possibly also in H~{\sc ii} regions. 

\section{Nebular conditions as probed by recombination lines and continua}

The bi-abundance model proposed by Liu et al. (2000) assumes that most of the
observed flux of ORLs are emitted by cold, H-deficient inclusions embedded in
the nebula and predicts that, for a given nebula, $T_{\rm e}$([O~{\sc iii}])
$\gtrsim$ $T_{\rm e}$(H~{\sc i} BJ) $\gtrsim$ $T_{\rm e}$(He~{\sc i}) $\gtrsim$
$T_{\rm e}$(O~{\sc ii}) (Liu 2003; P\'{e}quignot et al. 2003). The predictions
are supported and strengthened by new observations and diagnostics developed
since the last IAU symposium on PNe.  Table~1 compares $T_{\rm e}$'s
and $N_{\rm e}$'s deduced from CELs and from ORLs/continua for PNe with adf 
$> 4.5$.  Two peculiar PNe, M\,2-24 (adf = 17, Zhang \& Liu 2003) and M\,3-27 
(adf = 5.5, Wesson, Liu \& Barlow 2005) have been excluded from the list.

\begin{table}
  \begin{center}
  \caption{Comparison of $T_{\rm e}$'s and $N_{\rm e}$'s deduced from CELs and from ORLs/continua}
  \label{tab:te}
  \begin{tabular}{lccccccc}\hline
    Nebula & adf(O$^{++}$/H$^+$) & $T_{\rm e}$([O~{\sc iii}]) & $\log N_{\rm e}$(CELs) &  $T_{\rm e}$(BJ) & $\log N_{\rm e}$(BD) & $T_{\rm e}$(He~{\sc i}) & $T_{\rm e}$(O~{\sc ii}) \\
           &                     &                        (K) &            (cm$^{-3}$) &              (K) &            (cm$^{-3}$) &                     (K) &                     (K) \\\hline
NGC 7009  & 4.7 & 9980 &3.6 & 7200 &3.8 & 5040 &  420 \\
H 1-41    & 5.1 & 9800 &3.0 & 4500 &    & 2930 & $<$288 \\
NGC 2440  & 5.4 &16150 &3.8 &14000 &    &      & $<$288 \\
Vy 1-2    & 6.2 &10400 &3.4 & 6630 &    & 4430 & 3450 \\
IC 4699   & 6.2 &11720 &3.3 &12000 &    & 2460 & $<$288 \\
NGC 6439  & 6.2 &10360 &3.7 & 9900 &5.5 & 4900 &  851 \\
M 3-33    & 6.6 &10380 &3.3 & 5900 &    & 5020 & 1465 \\
M 2-36    & 6.9 & 8380 &3.6 & 6000 &3.8 & 2790 &  520 \\
IC 2003   & 7.3 &12650 &3.5 &11000 &3.0 & 5600 & $<$288 \\
NGC 6153  & 9.2 & 9120 &3.5 & 6000 &3.8 & 3350 &  350 \\
LMC N66   &11.0 &18150 &3.6 &      &    &      &      \\
DdDm 1    &11.8 &12300 &3.6 &11400 &3.8 & 3500 &      \\
Vy 2-2    &11.8 &13910 &4.2 & 9300 &    & 1890 & 1260 \\
NGC 2022  &16.0 &15000 &3.2 &13200 &    &15900 & $<$288 \\
NGC 40    &17.8 &10600 &3.1 & 7000 &3.2 &10240 &      \\
M 1-42    &22.0 & 9220 &3.1 & 4000 &3.7 & 2260 & $<$288 \\
NGC 1501  &31.7 &11100 &3.0 & 9400 &    &      &      \\
Hf 2-2    &71.2 & 8820 &3.0 & 1000 &2.6 &  940 &  630 \\\hline
  \end{tabular}
\begin{description}
\item {\em References:}\, NGC7009: Liu et al.(1995); H\,1-41, IC\,4699, 
NGC\,6439, M\,3-33: Wang \& Liu (in preparation); NGC\,2440, NGC\,2022, 
LMC\,N\,66: Tsamis et al. (2004); Vy\,1-2, IC\,2003, DdDm\,1, Vy\,2-2: Wesson, Liu
\& Barlow (2005); M\,2-36, M\,1-42: Liu et al. (2001); NGC\,6153: Liu et al.
(2000); NGC\,40: Liu et al. (2004b); NGC\,1501: Ercolano et al. (2004); 
Hf\,2-2: Liu et al. (2006).
\end{description}
 \end{center}
\end{table}

\begin{figure}
\centering
\resizebox{9.6cm}{!}{\includegraphics{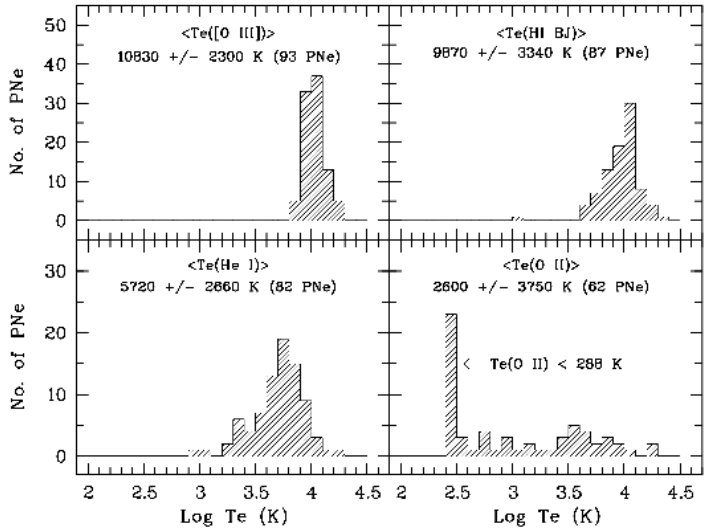}}
\caption[]{Distributions of $T_{\rm e}$ derived from the a) [O~{\sc
iii}] nebular to auroral line ratio; b) H~{\sc i} Balmer jump; c)
He~{\sc i} line ratios (mostly $\lambda$7281/$\lambda$6678); and d) O~{\sc
ii} $\lambda$4089/$\lambda$4649 line ratio. For the distribution in
each panel, the average $T_{\rm e}$ and its standard deviation are labelled.}
\end{figure}

\subsection{H~{\sc i} recombination spectrum}

For nearly all PNe surveyed, $T_{\rm e}$(BJ) has been determined from the ratio
of Balmer jump at 3646\,{\AA} to H\,11 $\lambda$3770 of the H~{\sc i}
recombination spectrum. Whereas the Balmer discontinuity is sensitive to
$T_{\rm e}$, intensities of high-order Balmer lines ($n \gtrsim 20$) relative
to a lower one (e.g. H\,11) increase with $N_{\rm e}$ and can thus be used to
deduce $N_{\rm e}$ under which H~{\sc i} lines are emitted. Zhang et al. (2004)
deduced values of $T_{\rm e}$ and $N_{\rm e}$ from the Balmer jump and
decrement, $T_{\rm e}$(BJ) and $N_{\rm e}$(BD), for a sample of 48 PNe.  Apart
from the well established relation that $T_{\rm e}$([O~{\sc iii}]) $\gtrsim
T_{\rm e}$(BJ), they also find that in general $N_{\rm e}$(BD) $\gtrsim N_{\rm
e}$(CELs), the average density deduced from optical CELs (c.f.  Table~1).
In addition, they find that for many PNe, $T_{\rm e}$([O~{\sc iii}]) deduced
from the ($\lambda$4959 + $\lambda$5007)/$\lambda$4363 ratio is lower than that
yielded by the ($\lambda$4959 + $\lambda$5007)/(52$\mu$m + 88$\mu$m) ratio,
which is in contradiction with the predictions of $T_{\rm e}$ fluctuations but
can be explained by invoking modest $N_{\rm e}$ inhomogeneities.

\subsection{He~{\sc i} recombination spectrum}

Liu (2003) used the He~{\sc i} $\lambda$5876/$\lambda$4471 and
$\lambda$6678/$\lambda$4471 ratios to deduce the average $T_{\rm e}$ of He~{\sc
i} ORLs, $T_{\rm e}$(He~{\sc i}).  Zhang et al.  (2005a) proposed to use
another ratio $\lambda$7281/$\lambda$6678 and analyzed 48 PNe.  The latter has
the advantages of being less sensitive to uncertainties in flux calibration,
reddening corrections and to optical depth effects of the metastable 2s\,$^3$S
level, although the lines involved are weaker. The results show that $T_{\rm
e}$(He~{\sc i}) $\lesssim$ $T_{\rm e}$(BJ), in line with the predictions of the
bi-abundance model, but in stark contradiction with the scenario of $T_{\rm e}$
fluctuations which predicts that $T_{\rm e}$(He~{\sc i}
$\lambda$7281/$\lambda$6678) $\gtrsim$ $T_{\rm e}$(BJ).  Another potential
diagnostic to determine $T_{\rm e}$(He~{\sc i}), yet to be fully exploited, is
the weak He~{\sc i} discontinuity at 3421\,{\AA} produced by recombinations of
He$^+$ to the 2p\,$^3$P$^{\rm o}$ level (Liu \& Danziger 1993; Zhang et al.
2005b).  Recombinations of He$^{++}$ to the $n = 5$ level also produce a
discontinuity at 5694\,{\AA} in the optical, which is however only observable
in very high excitation PNe where helium is mostly doubly ionized (Liu \&
Danziger 1993).

\subsection{O~{\sc ii} ORLs}

In their analysis of M\,1-42 and M\,2-36, Liu et al. (2001) noticed that while
the relative intensities of O~{\sc ii} ORLs observed in those two PNe and in
NGC\,7009 (Liu et al. 1995) and NGC\,6153 (Liu et al. 2000) agree well with the
predictions of recombination theory, there are discrepancies.  For example,
relative to the strongest 4f -- 4d transition 4f\,G[5]$^{\rm o}_{11/2}$ --
3d\,$^4$F$_{9/2}$ $\lambda$4089, the intensity of the strongest 3p -- 3s
transition 3p\,$^4$D$^{\rm o}_{7/2}$ -- 3s\,$^4$P$_{5/2}$ $\lambda$4649 appears
to be too weak by $\sim 40$\% compared to the predicted value calculated at
$T_{\rm e}$(BJ).  Later it became clear that the discrepancy was caused by the
fact that the O~{\sc ii} ORLs arise from plasma of $T_{\rm e}$'s far lower than
$T_{\rm e}$(BJ) (Liu 2003). By comparing the observed
$\lambda$4089/$\lambda$4649 ratio with the theoretical value as a function of
$T_{\rm e}$ calculated down to 288~K, the average $T_{\rm e}$'s under which
O~{\sc ii} lines are emitted have been determined for a number of PNe (Tsamis
et al.  2004; Liu et al. 2004b and Wesson et al. 2005). In general, $T_{\rm
e}$(O~{\sc ii}) $\sim 1000$~K (Table~1).  Fig.\,3 shows histograms of $T_{\rm
e}$([O~{\sc iii}]), $T_{\rm e}$(BJ), $T_{\rm e}$(He~{\sc i}) and $T_{\rm
e}$(O~{\sc ii}).  While $T_{\rm e}$(O~{\sc ii}) scatters over a wide range of
values, presumably due to measurement uncertainties given the weakness of the
lines, the diagram clearly shows that $T_{\rm e}$(O~{\sc ii}) $\lesssim T_{\rm
e}$(He~{\sc i}) $\lesssim T_{\rm e}$(BJ) $\lesssim T_{\rm e}$([O~{\sc iii}]),
as predicted by the bi-abundance model (Liu et al. 2000). Note that the width
of the $T_{\rm e}$ distribution increases from $T_{\rm e}$([O~{\sc iii}]),
through $T_{\rm e}$(BJ) and $T_{\rm e}$(He~{\sc i}) to $T_{\rm e}$(O~{\sc ii}).
Parts of this increase are clearly caused by observational uncertainties as the
measurements become increasingly difficult. Alternatively, it may well be that
the relative importance of the postulated cold H-deficient inclusions and the
main nebula in emitting ORLs varies from object to object; this contributes to
the large scatter observed in $T_{\rm e}$(He~{\sc i}) and particularly in
$T_{\rm e}$(O~{\sc ii}).  This is consistent with the loose correlation
observed between adf and $T_{\rm e}$([O~{\sc iii}])$- T_{\rm e}$(BJ) (Fig.\,1). 

Applying the same technique to the well known H-deficient knots in Abell\,30,
Wesson et al. (2003) showed that the prominent O~{\sc ii} ORLs emitted by
those knots arise from plasma of $T_{\rm e} \sim 500$ -- 2000~K.  Note that the
knots are O-rich, in contradiction to the prediction that they originate from
ejecta of a late thermal pulse of the helium flash of a {\em single}\,
star evolution and should therefore be C-rich (Iben et al. 1983).

Several PNe in our sample show O~{\sc ii} $\lambda$4089/$\lambda$4649 ratios
higher than the predicted value of 0.41 at 288~K, the lowest $T_{\rm e}$ for
which the effective recombination coefficients are available. Tsamis et al.
(2004) detected a weak feature at 4116\,{\AA} in 3 PNe (NGC\,3242, 5882 and
6818) and attributed it to Si~{\sc iv} 4s\,$^2$S$_{1/2}$ -- 4p\,$^2$P$^{\rm
o}_{1/2}$ $\lambda$4116.1. The other component of the multiplet, $J = 1/2 -
3/2$ at 4088.8\,{\AA}, blends with O~{\sc ii} $\lambda$4089.3. Assuming Si~{\sc
iv} $\lambda$4089/$\lambda$4116 = 2, they corrected the measured O~{\sc ii}
$\lambda$4089 flux for the presence of Si~{\sc iv} when calculating $T_{\rm
e}$(O~{\sc ii}). For NGC\,7009 listed in Table~1, our unpublished spectrum
reveals a feature at 4116\,{\AA}. If we attribute it to Si~{\sc iv} and apply
similar corrections, then $T_{\rm e}$(O~{\sc ii}) increases from the listed
value of 420~K to 1310~K. On the other hand, the spectrum of another PN in
Table~1, Vy\,2-2, shows a feature at 4116\,{\AA} with an intensity that is
higher than even that of the $\lambda$4089 feature. In view of the facts that
O$^{2+}$/H$^+$ derived from the $\lambda$4089 feature, assuming it is entirely
due to O~{\sc ii} $\lambda$4089, is in good agreement with values derived from
other O~{\sc ii} lines (Wesson, Liu \& Barlow 2005), and that Si~{\sc iv}
$\lambda$4089/$\lambda$4116 = 2, it is unlikely that the $\lambda$4116 feature
detected in Vy\,2-2 is entirely due to Si~{\sc iv}, even if Si~{\sc iv} is
present in this nebula. We were unable to find a more plausible assignment for
the $\lambda$4116 feature.

Detailed analyses (Liu 2003; Tsamis et al. 2003) revealed that the relative
intensities of O~{\sc ii} ORLs within a given multiplet deviate from the
predictions of the recombination theory that has hitherto tacitly assumed that
the ground fine-structure levels of recombining O$^{++}$ ions,
2p$^2$\,$^3$P$_{0,1,2}$, are thermalized and populated according to the
statistical weights (Storey 1994; Liu et al. 2000).  The effects provide a
means of determining the density under which the lines are emitted, and
consequently the total mass of ionized gas required to reproduce their observed
fluxes. An empirical calibration of the intensity of the 3p\,$^4$D$^{\rm
o}_{7/2}$ -- 3s\,$^4$P$_{5/2}$ $\lambda$4649 transition relative to the total
intensity of the whole multiplet as a function of {\em forbidden line}\,
$N_{\rm e}$ was presented by Ruiz et al. (2003).  {\it Ab initio}\,
calculations of effective recombination coefficients that take into account the
populations of the individual fine-structure levels of the recombining ion as a
function of $N_{\rm e}$ have been carried out and preliminary results reported
by Bastin \& Storey (2005; c.f. also the contribution by the same authors in
the volume). Applications of the data to observations yield densities that are
higher than the diffuse gas, as one would expect if the cold high metallicity
plasma that is postulated to be responsible for most of the observed emission
of ORLs originates from evaporation of H-deficient condensations embedded in
the nebula. Analysis also shows that the amount of metal deposited in those
H-deficient clumps is substantial and comparable to that in the ``normal''
component (Liu et al. 2006).

In summary, a self-consistent picture has emerged that points to the presence
of a new component of cold plasma that is highly enriched in helium and heavy
elements and probably in the form of H-deficient inclusions embedded in the
nebula. Its existence provides a natural solution to the long-standing
dichotomy of nebular plasma diagnostics and abundance determinations using ORLs
on the one hand and CELs on the other. The study has demonstrated that we have
much to learn from those weak ORLs that are only revealed by deep high
resolution spectroscopy. The diagnostic tools developed from the analysis of
the O~{\sc ii} spectrum as outlined above can be easily extended to
recombination spectra of other heavy element ions, such as C~{\sc ii}, N~{\sc
ii} and Ne~{\sc ii}, though the observations would be even more demanding as
the lines become even fainter.  Indeed, analyses of available data on C~{\sc
ii} and N~{\sc ii} ORLs, albeit sparse, yield corroborative evidence that they
arise from plasma of $T_{\rm e} \sim 1000$~K, as do the O~{\sc ii} ORLs.
 
Several scenarios for the possible origins of the postulated H-deficient
inclusions have been proposed, including the ``born-again'' scenario (Iben et
al. 1983), evaporating planetesimals (Liu 2003, 2005) and novae (Wesson et al.
2003). Further studies are clearly needed to discriminate between these
possibilities.


\begin{thebibliography}{}

\bibitem[Barker (1987)]{barker87}{Barker, T.} 1987, \textit{ApJ} 322, 922

\bibitem[Barlow et al. (2003)]{barlow03}{Barlow, M. J., et al.} 
        2003, in: S. Kwok, M. Dopita \& R. Sutherland (eds.), \textit{Planetary Nebulae: Their Evolution and Role in the Universe} (San Francisco: ASP), p.373 

\bibitem[Bastin, R. \& Storey (2005)]{bastin05}{Bastin, R. \& Storey, P. J.} 2005, in: R. Szczerba, G. Stasinska \& S. K. Gorny (eds.), \textit{Planetary Nebulae as Astronomical Tools} (Springer, New York), AIP Conf. Proc., Vol.\,804, p.63 

\bibitem[Ercolano \etal (2004)]{ercolano04}{Ercolano, B., Wesson, R., Zhang, Y., et al.} 
         2004, \textit{MNRAS} 354, 558

\bibitem[Esteban \etal (2002)]{esteban02}{Esteban, C., Peimbert, M., Torres-Peimbert, S., 
        \& Rodr\'{i}guez, M.} 2002, \textit{ApJ} 581, 241

\bibitem[Esteban \etal (2004)]{esteban04}{Esteban, C., Peimbert, M., Garc\'{i}a-Rojas, J., et al.} 
        2004, \textit{MNRAS} 355, 229

\bibitem[Garc\'{i}a-Rojas \etal (2004)]{garcia04}{Garc\'{i}a-Rojas, J., Esteban, C., Peimbert, M., et al.} 
        2004, \textit{ApJS} 153, 501

\bibitem[Garc\'{i}a-Rojas \etal (2005)]{garcia05}{Garc\'{i}a-Rojas, J., Esteban, C., Peimbert, A., et al.} 
        2005, \textit{MNRAS} 362, 301

\bibitem[Garc\'{i}a-Rojas \etal (2006)]{garcia06}{Garc\'{i}a-Rojas, J., Esteban, C., Peimbert, M., et al.} 
        2006, \textit{MNRAS} 368, 253

\bibitem[Garnett \& Dinerstein (2001a)]{garnett01a}{Garnett, D. \& Dinerstein, H. L.} 2001a, \textit{ApJ} 558, 145

\bibitem[Garnett \& Dinerstein (2001b)]{garnett01b}{Garnett, D. \& Dinerstein, H. L.} 2001b, \textit{RMxAA (Conf. Ser.)} 10, 13

\bibitem[Iben, Kaler \& Truran (1983)]{iben83}{Iben, I., Kaler, J. B. \& Truran J. W.} 1983, \textit{ApJ} 264, 605

\bibitem[Liu (2003)]{liu03}{Liu, X.-W.} 2003, in: S. Kwok, M. Dopita \& R. Sutherland (eds.), \textit{Planetary Nebulae: Their Evolution and Role in the Universe} (San Francisco: ASP), p.339 

\bibitem[Liu (2005)]{liu05}{Liu, X.-W.} 2005, in: J. Walsh, L. Stanghellini \& N. Douglas (eds.), \textit{Planetary Nebulae beyond the Milky Way} (Berlin: Springer-Verlag), p.169

\bibitem[Liu \etal (2006)]{liu06}{Liu, X.-W., Barlow, M. J., Zhang, Y., et al.} 
        2006, \textit{MNRAS} in press (astroph/0603215) 

\bibitem[Liu \& Danziger (1993)]{liu93}{Liu, X.-W. \& Danziger, I. J.} 1993, \textit{MNRAS} 263, 256

\bibitem[Liu \etal (2004a)]{liuy04a}{Liu, Y., Liu, X.-W., Luo, S.-G. \& Barlow, M. J.} 2004a, \textit{MNRAS} 353, 1231

\bibitem[Liu \etal (2004b)]{liuy04b}{Liu, Y., Liu, X.-W., Barlow, M. J. \& Luo, S.-G.} 2004b, \textit{MNRAS} 353, 1251

\bibitem[Liu \etal (2001)]{liu01}{Liu, X.-W., Luo, S.-G., Barlow, M. J., Danziger, I. J. \& Storey, P. J.} 2001, \textit{MNRAS} 327, 141

\bibitem[Liu \etal (2001)]{liu01}{Liu, X.-W., Storey, P. J., Barlow, M. J. \& Clegg, R. E. S.} 2001, \textit{MNRAS} 272, 369

\bibitem[Lodders (2003)]{lodders03}{Lodders, K.} 2003,  \textit{ApJ} 591, 1220

\bibitem[Luo \& Liu (2003)]{luo03}{Luo, S.-G. \& Liu, X.-W.} 2003, in: S. Kwok, M. Dopita \& R. Sutherland (eds.), \textit{Planetary Nebulae: Their Evolution and Role in the Universe} (San Francisco: ASP), p.393 

\bibitem[Luo \etal (2001)]{luo01}{Luo, S.-G., Liu, X.-W. \& Barlow, M. J.} 2001, \textit{MNRAS} 326, 1049

\bibitem[Peimbert (1967)]{peimbert67}{Peimbert, M.} 1967, \textit{ApJ} 150, 825

\bibitem[Peimbert (1971)]{peimbert71}{Peimbert, M.} 1971, \textit{Bol. Obs. Tonantzintla Tacubaya} 6, 29

\bibitem[Peimbert (2003)]{peimbert03}{Peimbert, M., Peimbert, A., Ruiz, M. T. \& Esteban, C.} 2004, \textit{ApJS} 150, 431

\bibitem[Peimbert \etal (2004)]{peimbert04}{Peimbert, A.} 2003, \textit{ApJ} 584, 735

\bibitem[P\'{e}quignot et al. (2003)]{pequignot03}{P\'{e}quignot, D. et al.} 
        2003, in: S. Kwok, M. Dopita \& R. Sutherland (eds.), \textit{Planetary Nebulae: Their Evolution and Role in the Universe} (San Francisco: ASP), p.347 

\bibitem[Robertson-Tessi \& Garnett (2005)]{garnett05}{Robertson-Tessi, M. \& Garnett, D.} 2005, \textit{APJS} 157, 371

\bibitem[Rubin (1989)]{rubin89}{Rubin, R. H.} 1989, \textit{ApJS} 69, 897

\bibitem[Rubin et al. (2002)]{rubin02}{Rubin, R. H., et al.} 
        2002, \textit{MNRAS} 334, 777
 
\bibitem[Ruiz \etal (2003)]{ruiz03}{Ruiz, M. T., Peimbert, A., Peimbert, M. \& Esteban, C.} 2003, \textit{ApJ} 595, 247

\bibitem[Sharpee, Baldwin \& Williams (2004)]{sharpee04}{Sharpee, B., Baldwin, J. A. \& Williams, R.} 2004, \textit{ApJ} 615, 323

\bibitem[Shen, Liu \& Danziger (2003)]{shen03}{Shen, Z.-X., Liu, X.-W. \& Danziger, I. J.} 2003, \textit{A\&A} 422, 563

\bibitem[Storey (1994)]{storey94}{Storey, P. J.} 1994, \textit{A\&A} 282, 999

\bibitem[Tsamis \etal (2003a)]{tsamis03a}{Tsamis, Y. G., Barlow, M. J., Liu, X.-W., et al.} 
        2003a, \textit{MNRAS} 338, 186

\bibitem[Tsamis \etal (2003b)]{tsamis03b}{Tsamis, Y. G., Barlow, M. J., Liu, X.-W., et al.} 
        2003b, \textit{MNRAS} 345, 186

\bibitem[Tsamis \etal (2004)]{tsamis04}{Tsamis, Y. G., Barlow, M. J., Liu, X.-W., et al.} 
        2004, \textit{MNRAS} 353, 953

\bibitem[Viegas \& Clegg (1994)]{viegas94}{Viegas, S. \& Clegg, R. E. S.} 1994, \textit{MNRAS} 271, 993

\bibitem[Wesson, Liu \& Barlow (2003)]{wesson03}{Wesson, R., Liu, X.-W. \& Barlow, M. J.} 2003, \textit{MNRAS} 340, 253

\bibitem[Wesson \& Liu (2004)]{wesson04}{Wesson, R. \& Liu, X.-W.} 2004, \textit{MNRAS} 351, 1026

\bibitem[Wesson, Liu \& Barlow (2005)]{wesson05}{Wesson, R., Liu, X.-W. \& Barlow, M. J.} 2005, \textit{MNRAS} 362, 424

\bibitem[Zhang \& Liu (2003)]{zhang03}{Zhang, Y. \& Liu, X.-W.} 2003, \textit{A\&A} 404, 545

\bibitem[Zhang \etal (2004)]{zhang04}{Zhang, Y., Liu, X.-W., Wesson, R., et al.} 
        2004, \textit{MNRAS} 351, 935

\bibitem[Zhang \etal (2005a)]{zhang05a}{Zhang, Y., Liu, X.-W., Liu, Y. \& Rubin, R. H.} 2005a, \textit{MNRAS} 358, 457

\bibitem[Zhang \etal (2005b)]{zhang05b}{Zhang, Y., Liu, X.-W., Luo, S.-G., P\'{e}quignot, D. \& Barlow, M. J.} 2005b, \textit{A\&A} 442, 249

\bibitem[Zhang, Rubin \& Liu (2005)]{zhang05}{Zhang, Y., Rubin, R. H. \& Liu, X.-W.} 2005c, \textit{RevMexAA (Serie de Conf.)} 23, 15

\end{thebibliography}
\end{document}